\documentclass[twocolumn,amsmath,amssymb]{revtex4}

\usepackage{amsmath} %
\usepackage{graphicx}
\usepackage{amsfonts}
\usepackage{dcolumn}
\usepackage{bm}

\begin{document}
\title{An Efficient Self-optimized Sampling Method for Rare Events in Nonequilibrium Systems}

\author{Huijun Jiang}
\author{Mingfeng Pu}
\author{Zhonghuai Hou}
\thanks{Corresponding author. E-mail: hzhlj@ustc.edu.cn}

\affiliation{Department of Chemical Physics \& Hefei National Laboratory for Physical Sciences at Microscales, University of Science and Technology of China, Hefei, Anhui 230026, China}

\date{\today}

\begin{abstract}
Rare events such as nucleation processes are of ubiquitous importance in real systems. The most popular method for nonequilibrium systems, forward flux sampling (FFS), samples rare events by using interfaces to partition the whole transition process into sequence of steps along an order parameter connecting the initial and final states. FFS  usually suffers from two main difficulties: low computational efficiency due to bad interface locations and  even being not applicable when trapping into unknown intermediate metastable states. In the present work,  we propose an  approach to overcome these difficulties, by self-adaptively locating the  interfaces on the fly in an optimized manner. Contrary to the conventional FFS which set the interfaces with euqal distance of the order parameter, our approach determines the interfaces with equal transition probability which is shown to satisfy the optimization condition. This is done  by firstly running long local trajectories starting from the current interface $\lambda_i$ to get the conditional probability distribution $P_c(\lambda>\lambda_i|\lambda_i)$, and then determining $\lambda_{i+1}$  by equalling $P_c(\lambda_{i+1}|\lambda_i)$ to a give value $p_0$. With these optimized interfaces, FFS can be run in a much efficient way. In addition, our approach can conveniently find the intermediate metastable states by monitoring some special long trajectories that nither end at the initial state nor reach the next interface, the number of which will increase sharply from zero if such  metastable states are encountered. We apply our approach to a model two-state system  and a two-dimensional lattice gas Ising model. Our approach is shown to be much more efficient than the conventional FFS method without losing  accuracy, and it  can also well reproduce the  two-step nucleation scenario of the Ising model with easy identification of the intermidiate metastable state.
\end{abstract}


\maketitle
\section{Introduction}
Many important transition processes in real systems are rare events, including nucleation, protein folding, polymer translocation through nanopore, etc. Rare events are usually fluctuation driven barrier-crossing events, occurring with a very low probability, but may be followed by important consequence when they do happen.  To study rare events theoretically, one needs specific sampling methods to calculate the transition rate and to identify the transtion pathways.  For equilibrium systems where detailed-balance is satisfied and equilibrium distribution is known, many methods have been proposed in the literature  \cite{Chandler1987,Frenkel2002,JCP78002959,ACP95000381,PRE01026109,
JCP98001964,ARPC02000291,JCP03007762,JComP05000157,JCP04010880}, including
'reactive flux'  and 'path sampling' methods. The former methods aim to sample the reactive flux by firing typical reactive trajectories near the transition state, such as the Bennett-Chandler method \cite{Chandler1987,Frenkel2002}, history dependent Bennett-Chandler method \cite{JCP78002959}, effective positive flux formalism \cite{ACP95000381}, and so on. The latter ones try to sample the whole ensemble of transition paths from which many information can be drawn,  including the  Crooks and Chandler approach \cite{PRE01026109}, transition path sampling \cite{JCP98001964,ARPC02000291}, transition interface sampling \cite{JCP03007762,JComP05000157}, milestoning method \cite{JCP04010880}, etc. For nonequilibrium systems without detailed balance and known distribution, however, path-sampling methods are relatively rare.

\begin{figure}
\begin{center}
\includegraphics[width=0.9\columnwidth] {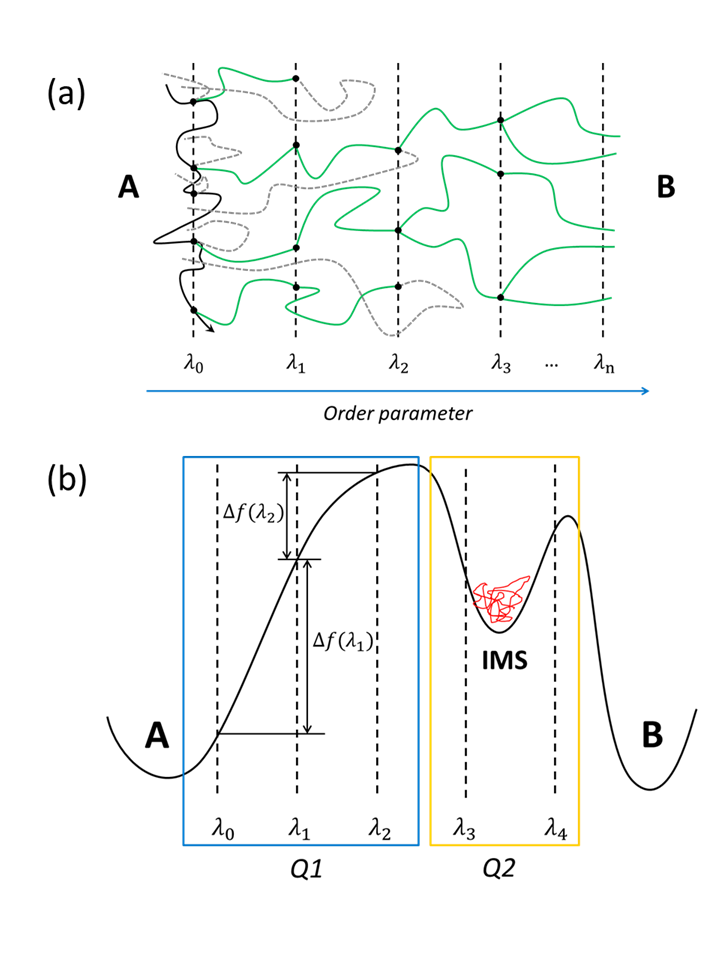}
\end{center}
\caption{ (a) Schematic illustrations for the FFS method. (b) Two main difficulties (Q1) and (Q2) encountered by FFS. (Q1): Setting interfaces with equal order parameter distance may result in large discrepancies in some  free-energy-like functional $f(\lambda)$ which causes low efficiency. (Q2): Possible hidden IMS traps trajectories which may make FFS unapplicable. See text for more details.} \label{fig:FFS}
\end{figure}

    Very recently, R. Allen et al. proposed the so-called  forward flux sampling (FFS) approach to study  rare events in nonequilibrium systems\cite{PRL05018104,JCP06024102,JPCM09463102}. As shown in Fig.\ref{fig:FFS}(a), FFS assumes an order parameter $\lambda$ which distinguishes the initial state $\mathbf{A}$ (where $\lambda<\lambda_0$) and the final state $\mathbf{B}$ (where $\lambda>\lambda_N$). A series of interfaces $\{\lambda_i,  i=1,...,N-1\}$ in between with $\lambda_{i+1}>\lambda_i$  are used to calculate the transition rate  and to sample the transition path ensemble. FFS generates partial trajectories between adjacent interfaces which are integrated forward in time only, not  requiring detailed balance.  The transition rate from $\mathbf{A}$ to $\mathbf{B}$ can be calculated by
\begin{equation}
 k_{AB}=\phi_0 \prod_{i=0}^{N-1}P(\lambda_{i+1}|\lambda_i),
\end{equation}\label{eq:kAB}
where $\phi_0$ is the effective forward flux leaving $\mathbf{A}$ and reaching interface $\lambda_0$, and $P(\lambda_{i+1}|\lambda_i)$ is the conditional probability that a trajectory coming from $\mathbf{A}$ crosses interface $\lambda_i$ for the first time and then reaches interface $\lambda_{i+1}$ before returning to $\mathbf{A}$. Typically, FFS processes as follows \cite{JCP06024102}:  (a) Run a long trajectory with time length $T$ starting from the initial state $\mathbf{A}$. This trajectory will cross $\lambda_0$ for $n_0$ times. Calculate the flux out from $\mathbf{A}$ by $\phi_0=n_0/T$. (b) Run $M_i$ trajectories with enough time length starting from the stored configurations at interface $\lambda_i$. There will be $n_i$ successful trajectories which cross interface $\lambda_{i+1}$  before returning to $\mathbf{A}$. Calculate the conditional probability by $P(\lambda_{i+1}|\lambda_i)=n_i/M_i$. Repeat this step utill the final state $\mathbf{B}$  is reached.  Due to its easy implementation, FFS has been widely used in  a variety of systems, e.g., the flipping of genetic toggle switch \cite{PRL05018104}, nucleation process \cite{PRL06065701,PRL07055702,NaC13001887}, polymer translocation \cite{JCP06024102,JCP08114903,JCP09044904}, protein conformational changes \cite{JCP06164904,JCP09225101}, to list just a few.

However, FFS also encounters several difficulties in its application. One  important issue is about the optimization of interface locations which will strongly affect computation efficiency. As illustrated in Fig.\ref{fig:FFS}(b) (Q1), interfaces in conventional FFS are usually set to be of equal order parameter distance, i.e., $\Delta \lambda=\lambda_{i+1}-\lambda_i$ is constant for all $i$. Assuming that there is an underlying free-energy-like function $f(\lambda)$ (  which may refer to the potential of mean force  for equilibrium systems or some action functional for nonequilibrium systems), the equal-distance interfaces may result in some time-consuming bottlenecks where the barrier $\Delta f(\lambda)$ between two adjacent interfaces is much larger than others,  for instance, $\Delta f(\lambda_1) >> \Delta f(\lambda_2)$ in  Fig.\ref{fig:FFS}(b). We note that an iterative FFS (IFFS) approach has been introduced to optimize the interface locations \cite{JCP08024115}, however,  it allways requires to run a complete FFS with un-optimized interfaces for the first time  which makes it even more inefficient than FFS.   Another issue is about  possible unknown intermediate metastable states (IMSs) hiding between the initial and final states wherein partial transition paths will be trapped  as shown in Fig.\ref{fig:FFS}(b) (Q2). The existence of such IMSs will make the conventional FFS (or IFFS) unapplicable. Thus, a new well designed approach which can optimize interface locations adaptively and search for the unknown IMSs automatically is very demanded.


In the paper, we propose a self-optimized FFS (SO-FFS) approach to overcome these two difficulties.
We first demonstrate that an optimized set of interfaces should have equal barrier height  $\Delta f(\lambda_i)=\max\{f(\lambda)-f(\lambda_i),\lambda_i\leq\lambda\leq\lambda_{i+1}\}$ rather than equal order-paramter distance $\Delta \lambda_i = \lambda_{i+1} - \lambda_i$. Consequently, the transition probability between adjacent interfaces should be nearly equal during the climbing stage. Therefore, we can sample the conditional transition probability $P_c(\lambda>\lambda_i|\lambda_i)$ from the current interface $\lambda_i$, by running local dynamics 'on the fly',  and the next interface $\lambda_{i+1}$ can be determined by equalling $P_c(\lambda=\lambda_{i+1}|\lambda_i)$ to a given value $p_0$. As long as a convergent $P_c(\lambda>\lambda_i|\lambda_i)$ is obtained, the method works efficiently over any profile of $f(\lambda)$.  In addition, the method also facilatates the identification of IMSs by monitoring special trajectories which end neither at the initial state nor at the next interface even for a sufficient long time. The number of these trajectories will increase sharply from zero around the IMSs. We apply our approach to a model two-state system and a lattice gas Ising model to demonstrate its efficiency as well as accuracy and its ability to find IMSs.

\section{The SO-FFS Approach} \label{SEC:SOFFS}

To begin,  we need to figure out the optimization condition for interface locations.  Given a potential-like function $f(\lambda)$, the transition probability from $\lambda_i$ to $\lambda_{i+1}$ should be proportional to $e^{-\Delta f(\lambda_i)}$, where $\Delta f(\lambda_i)=\max\{f(\lambda)-f(\lambda_i),\lambda_i\leq\lambda\leq\lambda_{i+1}\}$ is the barrier height in between. Hence  the relative  time cost $c_i$ for sampling transition between these two interfaces can  be estimated by $c_i\sim e^{\Delta f(\lambda_i)}$. The total time cost is then
\begin{equation}\label{eq:tc}
c=\sum_i c_i=\sum_i e^{\Delta f(\lambda_i)}.
\end{equation}
Minimization of Eq.(\ref{eq:tc}) with the constraint $\sum_i {\Delta f(\lambda_i)}=const$ (the total barrier height)  leads to $\Delta f(\lambda_i)e^{\Delta f(\lambda_i)}=\alpha$ for $(i=0,...,N-1)$, where $\alpha$ is a Lagrangian multiplier. Clearly, the optimization condition is
\begin{equation}\label{eq:oc}
\Delta f(\lambda_0)=\Delta f(\lambda_1)=...=\Delta f(\lambda_{N-1}),
\end{equation}
corresponding to equal transition probability between adjacent interfaces.

\begin{figure}
\begin{center}
\includegraphics[width=0.8\columnwidth] {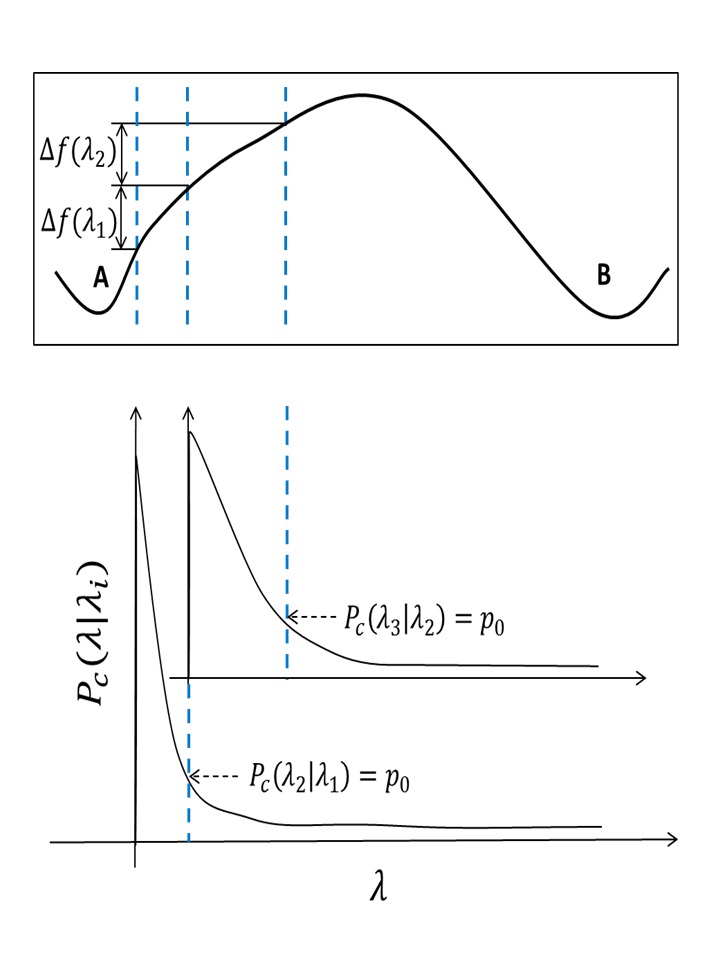}
\end{center}
\caption{ Schematic illustration of adaptive determination of new interfaces in SO-FFS. $P_c(\lambda|\lambda_i)$ denotes the conditional probability distribution sampled by local dynamics starting from configurations stored at interface $\lambda_i$. The next interface $\lambda_{i+1}$ is determined via $P_c(\lambda_{i+1}|\lambda_i)=p_0$ as shown in the bottom panel. For this choice of interface locations, the barrier heights $\Delta f(\lambda_i)$ are nearly the same as  shown in the top panel. } \label{fig:OC}
\end{figure}

Such a fact actually provides a simple but efficient way to sample the rare events with any unknown profile of $f(\lambda)$. We can perform local dynamics simulation to get the conditional probability distribution  $P_c(\lambda|\lambda_i)$ with $\lambda>\lambda_i$. Such distribution usually has an exponential-decaying form governed by large deviation law.
The next interface $\lambda_{i+1}$  is located at where $P_c(\lambda_{i+1}|\lambda_i)=p_0$, where $p_0$ is  a given fixed value, such that new interfaces can be obtained successively. As shown in Fig.\ref{fig:OC}, for instance, $\lambda_3$ can be obtained from $P_c(\lambda|\lambda_2)$ the same way as getting $\lambda_2$ from $P_c(\lambda|\lambda_1)$. Notice that the interfaces determined in this way are already self-optimized approximately.  With these interfaces, one can then canclulate the exact transition probabilities $P(\lambda_{i+1}|\lambda_i)$  in the same manner as in FFS.

\begin{figure}
\begin{center}
\includegraphics[width=1.1\columnwidth] {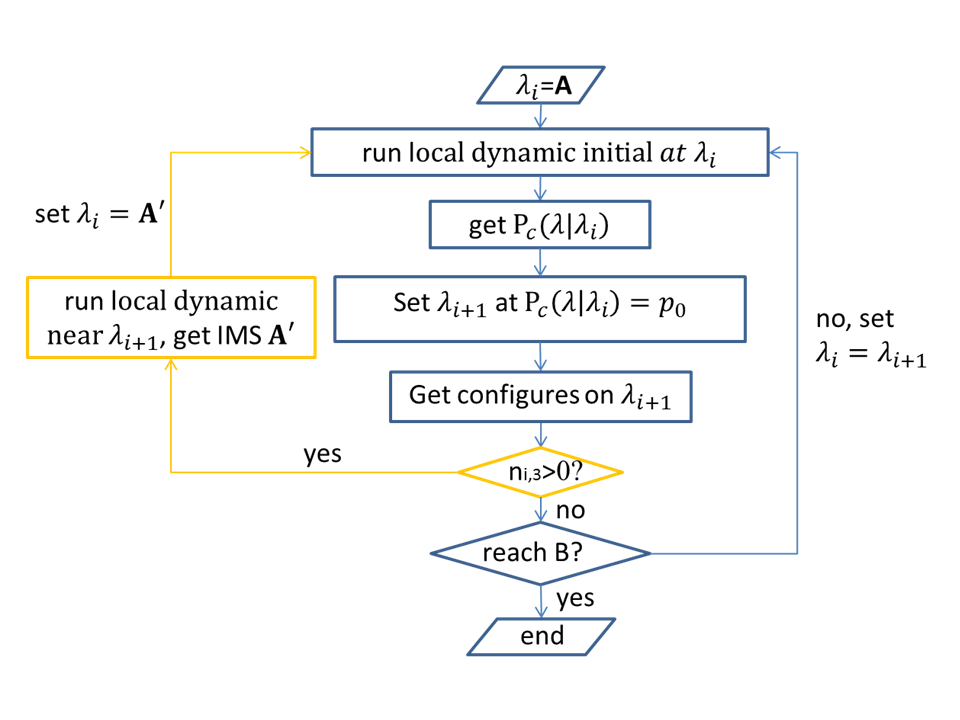}
\end{center}
\caption{ Simulation scheme of  the SO-FFS approach.} \label{fig:SOFFS}
\end{figure}

A detailed procedure is shown in Fig.\ref{fig:SOFFS} (blue lines):

(a) Run a long enough trajectory with time length $T$ starting from $\mathbf{A}$ until the conditional probability $P_c(\lambda|\mathbf{A})$  converges. Interface $\lambda_0$ is located at where the conditional probability is of a fixed value $P_c(\lambda_0|\mathbf{A})=p_0$. Store the configurations that cross $\lambda_0$ and count its number $n_0$. Calculate $\phi_0$ in the same way as in FFS by $n_0/T$.

(b) Run trajectories with fixed length $T_1$ starting from random-chosen configurations at interface $\lambda_i$ until the conditional probability $P_c(\lambda|\lambda_i)$ for $\lambda\geq\lambda_i$ converges. Interface $\lambda_{i+1}$ is located at where the conditional probability $P_c(\lambda|\lambda_i)$ takes value $p_0$. Store the configurations crossing $\lambda_{i+1}$ from left.

(c)  Calculate the {\it exact} conditional probability by $P(\lambda_{i+1}|\lambda_i)=n_i^1/M_i$, where $n_i^1$ is the number of successful trajectories which cross interface $\lambda_{i+1}$ before returning to $\mathbf{A}$ and $M_i$ is the total number of trajectories.  Notice that, trajectories which reach neither interface $\lambda_{i+1}$ nor the initial state $\mathbf{A}$ during $T_1$ should be further run till ending at one of the two states.

(d) Repeat step (b) and (c) till the final state $\mathbf{B}$ is reached.

In short words, the SO-FFS approach basically can be separated into two parts:  Determining the next interface self-adaptively by local dynamics  and running typical FFS by the interfaces obtained.

In practice, conditional probability $P_c(\lambda|\lambda_i)$ can also be replaced by the local distribution $\rho_c(\lambda|\lambda_i)$ which is much easier to be sampled. This further improves the  computation efficiency. The difference between $\rho_c(\lambda|\lambda_i)$ and $P_c(\lambda|\lambda_i)$ lies in that trajectories that cross a given $\lambda$ for multiple times contribute many times to the former but only once to the latter. Surely $\rho_c$ only appximates $P_c$, nevertheless, we find it already works very well. In addition, one may also use cumulant distribution $\int_{\lambda_i}^{\lambda}\rho_c(\lambda|\lambda_i)$ instead of $\rho_c$ itself since the cumulant distribution converges more rapidly. According to the optimization condition Eq.(\ref{eq:oc}), the next interface $\lambda_{i+1}$ can be located approximately by $\int_{\lambda_i}^{\lambda}\rho_c(\lambda|\lambda_i)=\rho_0$, where $\rho_0$ is a given value for the cumulant distribution.


Another problem encountered in application of FFS is that unknown IMMs may exist in complex systems, such as the multistep nucleation process found in Ising model \cite{PRL06065701}. When such an IMS presents, trajectories will hardly return to $\mathbf{A}$ for  $\lambda_i>\lambda_{\rm IMS}$ and will be trapped in the attractive basin of the IMS, which makes the FFS much time-consuming or even unapplicable.  If no IMS exists between $\mathbf{A}$ and $\mathbf{B}$,  the trajectories used for sampling the conditional probability $P_c(\lambda_{i+1}|\lambda_i)$ end either at  $\lambda_{i+1}$ (whose number is $n_{i,1}$) or at $\mathbf{A}$ (number of which is $n_{i,2}$).
In the presence of an IMS,  however, trajectories can be trapped by it which neither reach   $\lambda_{i+1}$ nor go back to $\mathbf{A}$.  Consequently, the number $n_{i,3}$ of such special trajectories will increase sharply from zero if $\lambda$ bypasses the IMS, which can be used as a fingerprint of the presence of IMSs. As shown in Fig.\ref{fig:SOFFS} (yellow lines), few substeps can be added  to search possible IMSs:

(c1) Count $n_{i,3}$ the number of trajectories which end  neither at $\lambda_{i+1}$ nor at $\mathbf{A}$ for a sufficient long time. If $n_{i,3}$ is substantially than zero, run a long enough trajectory starting from a configuration picked up randomly from these trajectories till the local phase space density $\rho(\lambda)$ converges.

(c2) Locate the IMS $\mathbf{A}'$ at  $\lambda_{\rm IMS}$ where $\rho(\lambda=\lambda_{\rm IMS})$ reaches the maximal value. Replace $\mathbf{A}$ by the IMS $\mathbf{A}'$, repeat the steps (a)-(c).

\section{Applications}

In this section, we will apply the SOFFS to a typical double-well system  and  a two-step nucleation process of Ising model. We use the former to demonstrate the efficiency of SO-FFS and the latter to show the ability of finding IMSs.

\subsection{Efficiency Comparison}

\begin{figure}
\begin{center}
\includegraphics[width=0.8\columnwidth] {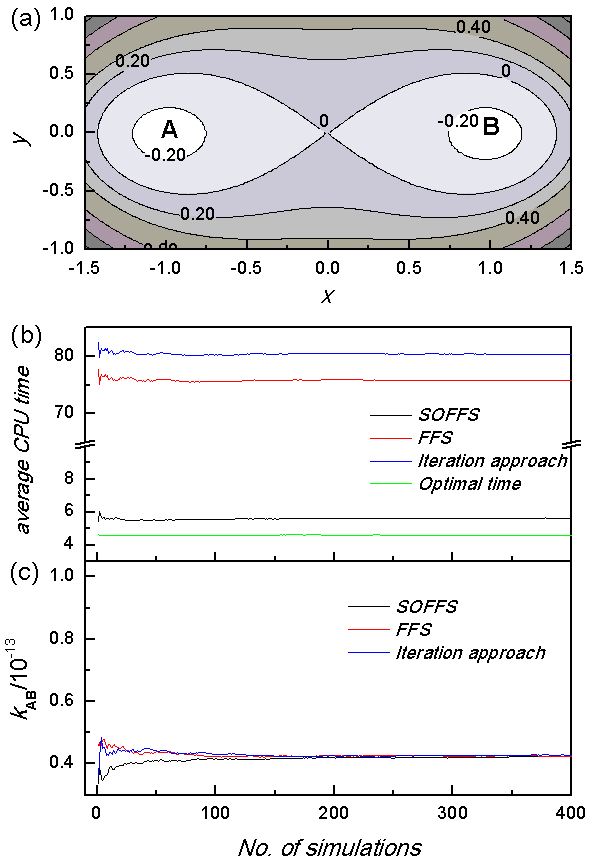}
\end{center}
\caption{(a) Potential landscape of the Maier-Stein model for $\beta=1$. (b) CPU time and (c) Average transition rate sampled by SO-FFS, FFS, and IFFS. $\beta=2$ for (b) and (c). } \label{fig:EC}
\end{figure}

We consider a typical double-well system, the Maier-Stein model, whose dynamics can be described as \cite{PRL93001783}:

\begin{equation}\label{eq:MS}
\begin{split}
\dot x (t) &= x   - ux^3  - \beta xy^2 + \sqrt {2D} \eta _x (t)   \\
\dot y (t) &=  y   - x^2 y + \sqrt {2D} \eta _y (t)
\end{split},
\end{equation}

\noindent where $\{\eta_x, \eta_y\}$ represent white noise with correlation $\langle\eta_u(t)\eta_v(t')\rangle=\delta_{u,v}\delta(t-t')$ and $D$ denotes the noise intensity.  The parameter $\beta$ reflects the conservation extent of the system. As shown in Fig.\ref{fig:EC}(a), for $\beta=1$, the drift field of the system can be viewed as a gradient of a potential  with two minima at $\mathbf{A}=(-1,0)$ and $\mathbf{B}=(1,0)$.  When small noise is present,  rare  transitions from $\mathbf{A}$ to $\mathbf{B}$ or vice versa are allowed. For $\beta\neq1$, $\mathbf{A}$ and $\mathbf{B}$ are still the asymptotic fixed points of the system, however, the system is not conserved and the system will not reach an 'equilibrium' state.
We use SO-FFS to study the nonequilibrium transition from $\mathbf{A}$ to $\mathbf{B}$ and compare its efficiency and accuracy with those of the FFS and IFFS.

The dynamical equation Eq.(\ref{eq:MS}) is integrated by Euler method with $D=0.01$ and time step $dt=0.01$. We fix $\beta=2$ to ensure the nonequilibrium feature of the transition process. The length of trajectory for SO-FFS is taken as $T_1=100dt$, and the threshold of cumulated distribution is $\rho_0=0.92$. The number of interfaces used in FFS and the IFFS is  the same as that obtained by SO-FFS. In FFS, the interfaces are set to divide equally the order parameter $\lambda$. A typical personal computer with a 3.0 GHz Interl(R) Core  CPU and 2 GB memory is used as the computation platform. The computation efficiency is measured by CPU time for sampling the transition. To get statistically reliable results, all the approaches are repeated for 400 times.

The results are shown in Fig.\ref{fig:EC}(b) and (c), where the CPU time and the calculated transition rate are shown, respectively. Clearly,  SO-FFS remarkably increases the computation efficiency compared FFS and IFFS. The average CPU time for SO-FFS is about $5.7s$, which is much smaller than  $75.6s$  for FFS and $80.2s$ for IFFS. In addition, SO-FFS nearly reaches the best efficiency (green line) which is obtained by using the optimal interfaces after 3 iterations of IFFS. Moreover,  besides its  high efficiency, SO-FFS can get the transition rate $k_{\mathbf{A}\mathbf{B}}$ accurately as well as shown in Fig.\ref{fig:EC}(c). In short, SO-FFS is demonstrated to be very efficient in sampling of rare events without losing of sampling accuracy.

\subsection{Searching for IMSs}

\begin{figure}
\begin{center}
\includegraphics[width=0.8\columnwidth] {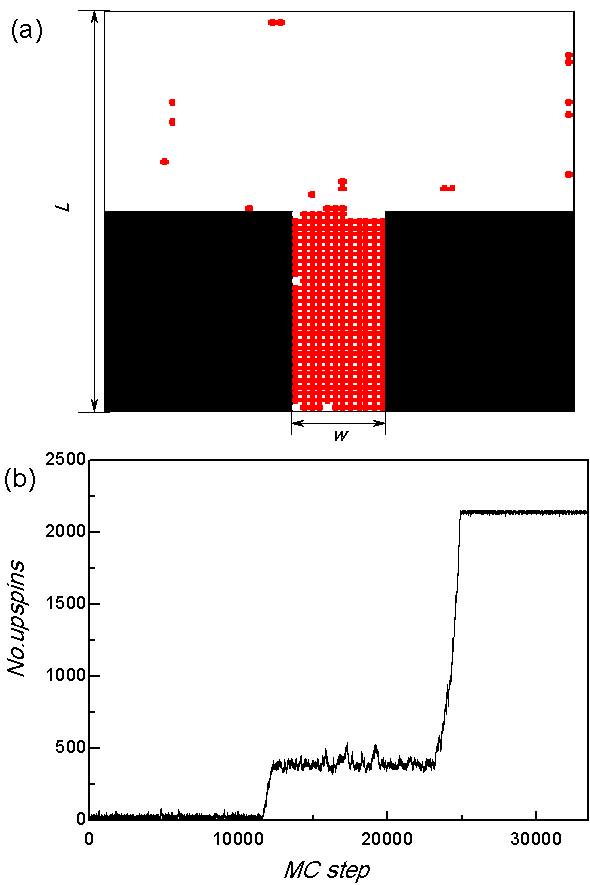}
\end{center}
\caption{(a) 2-dimensional lattice gas Ising model with a pore. The lattice size  $L\times L$ and the pore size  is $w\times L/2$. Red and white  sites are for up and down spins, respectively. The shown state corresponds to the intermediate stage in (b), where the first nucleation step inside the pore has just finished.  (b) A typical dynamical trajectory of the two-step nucleation for $L=60$ and $w=12$. } \label{fig:Ising}
\end{figure}

Here we consider the nucleation of a 2-dimensional lattice gas Ising model with pores which has been shown to be a two-step process \cite{PRL06065701}. As illustrated in Fig.\ref{fig:Ising}(a), the system is defined on a $L\times L$ square lattice and the pore has a simple $w\times L/2$ rectangular slit geometry with width $w$. Each site $i$ on the square lattice has a spin $s_i =\pm1$ associated with it. The Hamiltonian $E$ of the system consists of two parts \cite{PRL06065701}:

\begin{equation}\label{Eq:Ising}
E=J\sum_{ij} s_i s_j - h\sum_i s_i,
\end{equation}
where $J$ is the interaction strength and $h$ is an external magnetic field. The first summation runs  over all nearest-neighbor pairs of spins.  At low temperatures, the system is stable when all spins are down ($s=-1$) or up ($s=+1$) for $h=0$. When $h\neq0$, one of these two states will be metastable, and a nucleation process will occur if the system is initially at the metastable state. Here, we consider the nucleation process from the  spin-down initial state to the spin-up state by setting $h=0.05k_BT_0>0$ with $k_B$ the Boltzmann constant and $T_0$ the temperature. The system is studied by using Monte Carlo simulations \cite{Chandler1987}. If not otherwise stated, the spin interaction strength is fixed at $J=0.8k_BT_0$. As shown in Fig.\ref{fig:Ising}(b), a  two-step nucleation is found: the system firstly nucleates in the pore and then out of the pore \cite{PRL06065701}. By defining the total number of up spins as the order parameter $\lambda$, clearly, the system bypasses an IMS with $\lambda_{\rm IMS}$ approximately equal to the size of the pore.

\begin{figure}
\begin{center}
\includegraphics[width=0.8\columnwidth] {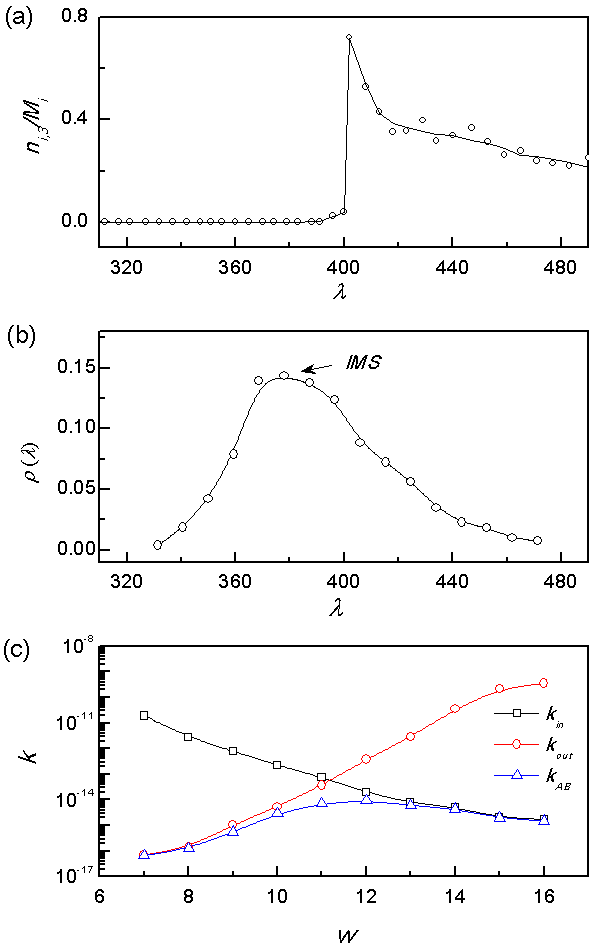}
\end{center}
\caption{(a) The fraction of sperical trajectories, $n_{i,3}/M_i$ as a function of the order parameter $\lambda$ which undergoes a sharp increase when passing the IMS. (b) The local  distribution $\rho(\lambda)$ near the IMS, which shows a clear-cut maximum corresponding to the IMS. (c) Nucelation rate as a function of the pore width $w$ sampled by SO-FFS.   $k_{in}$, $k_{out}$ and $k_{\mathbf{A}\mathbf{B}}$ denote the rate for the first, the second and the whole step, respectively. $L=60$ for all and $w=12$ for (a) and (b). } \label{fig:IMSS}
\end{figure}

We then sample the two-step nucleation by SO-FFS. As described in the last part of Section \ref{SEC:SOFFS}, we draw $n_{i,3}$(the number of special trajectories) as a function of $\lambda$ in Fig.\ref{fig:IMSS}(a). Apparently, a sharp jump is observed around $\lambda=400$, demonstrating an IMS nearby. We then sample the local phase space density $\rho(\lambda)$  starting from a configuration picked up randomly from these special trajectories. As shown in Fig.\ref{fig:IMSS}(b), $\rho(\lambda)$ shows a clear-cut maximum at $\lambda\simeq 380$  for $L=60$ and $w=12$, corresponding to the IMS. The value  $\lambda_{\rm IMS}=380$ is slightly larger than the pore size $\lambda=360$, which is reasonable since the  IMS describes the state wherein the nucleation in the pore has completed and the one out of the pore has just started. We have also used our  SO-FFS approach to sample the whole two-step nucleation process. To be specific, we have calculated the nucleation rates $k_{in}$  and $k_{out}$ inside and out of the pore respectively, as well as the whole nucleation rate  $k_{\mathbf{A}\mathbf{B}}=1/(1/k_{in}+1/k_{out})$), as functions of the pore  width $w$. The results are shown in Fig.\ref{fig:IMSS}(c), where $k_{in}$  decreases as $w$ increases, and $k_{out}$ increases with $w$. These competition between the two steps finally results in a non-monotonic dependence of the whole rate $k_{\mathbf{A}\mathbf{B}}$ on $w$, demonstrating an interesting type of size effect, in good agreements with the reported results\cite{PRL06065701}.

\section{Conclusion}
In summary, we have developed an efficient approach,  SO-FFS, to study rare events in nonequilibrium systems with  self-optimized computation efficiency and ability to find IMSs. Interfaces which divide the whole transition process into stages are determined adaptively by sampling local dynamics. SO-FFS ensures that the transition probability between adjacent interfaces are nearly the same, thus automatically works in an optimized manner. The method can also identify IMSs conveniently by monitoring some special trajectories that neither return back to the initial state nor reach the next interface. We show that our method is much more efficient that the conventional FFS without losing accuracy by applying it to a two-state model system. The ability of searching for IMSs is also demonstrated in a two-step nucleation process associted with lattice gas Ising model.  Although the two models considered here are relatively simple, the main physical idea is the same and  our SO-FFS approach can be easily applied to more complicated systems and will surely find wide applications.

\acknowledgments
This work is supported by National Science Foundation of China (21125313, 20933006, 91027012).

\bibliography{SOFFS}
\bibliographystyle{apsrev}

\end{document}